\documentclass[%
 reprint,
superscriptaddress,
 amsmath,amssymb,
 aps,
 prb,
]{revtex4-2}

\usepackage{graphicx}
\usepackage{dcolumn}
\usepackage{bm}
\usepackage{type1cm}
\usepackage{lettrine}

\usepackage{soul}
\usepackage{textcomp,gensymb}
\usepackage[dvipsnames]{xcolor}
\usepackage{tcolorbox}
\usepackage{svg}
\usepackage{siunitx}
\tcbset{width=0.9\textwidth,boxrule=0pt,colback=red,arc=0pt,auto outer arc,left=0pt,right=0pt,boxsep=5pt}

\usepackage{upgreek}
\usepackage{tikz}
\usepackage{makecell}

\begin{document}

\AtEndEnvironment{thebibliography}{
\bibitem{dataleeds}
        M.~Dearg, A.~J.~Caruana, C.~J.~Kinane, N.~Satchell, G.~Burnell, S.~Langridge, and C.~H.~Marrows, Research Data Leeds, https://doi.org/TBD, 2026.

\bibitem{dataisis}
        C.~H.~Marrows, C.~J.~Kinane, M.~Dearg, G.~Burnell, S.~Langridge, and N.Satchell, ISIS Neutron and Muon Source Data Journal, https://doi.org/10.5286/ISIS.E.RB1920567, 2019.
}


\title{Helical Magnetic Nature of $\beta$-Mn Structured CoZn Thin-Films Determined by a Bayesian Approach to Modeling Polarized Neutron Reflectivity Data}


\author{M. Dearg}
\altaffiliation[Present affiliation: ]{School of Physics, Engineering and Astronomy, University of York, York YO10 5DD, United Kingdom}
\affiliation{School of Physics and Astronomy, University of Leeds, Leeds LS2 9JT, United Kingdom}

\author{A. J. Caruana}
\email[Email: ]{andrew.caruana@stfc.ac.uk}
\affiliation{ISIS Neutron and Muon Source, STFC Rutherford Appleton Laboratory, Didcot, Oxon. OX11 0QX, United Kingdom}
\affiliation{Department of Physics, School of Science, Loughborough University, Loughborough LE11 3TU, United Kingdom}

\author{C. J. Kinane}
\affiliation{ISIS Neutron and Muon Source, STFC Rutherford Appleton Laboratory, Didcot, Oxon. OX11 0QX, United Kingdom}

\author{N. Satchell}
\altaffiliation[Present affiliation: ]{Department of Physics, Texas State University, San Marcos, Texas 78666, USA}
\affiliation{School of Physics and Astronomy, University of Leeds, Leeds LS2 9JT, United Kingdom}

\author{G. Burnell}
\affiliation{School of Physics and Astronomy, University of Leeds, Leeds LS2 9JT, United Kingdom}

\author{S. Langridge}
\affiliation{ISIS Neutron and Muon Source, STFC Rutherford Appleton Laboratory, Didcot, Oxon. OX11 0QX, United Kingdom}

\author{C. H. Marrows}
\email[Email: ]{c.h.marrows@leeds.ac.uk}
\affiliation{School of Physics and Astronomy, University of Leeds, Leeds LS2 9JT, United Kingdom}


\date{\today}

\begin{abstract}
Chiral magnetic materials can host skyrmions and hence are of interest for numerous spintronic applications including low-power storage. To reach this goal, room temperature operation and a thin film form are needed. We have studied the chiral magnetism in a $\beta$--Mn structured CoZn sputtered film, with a Curie temperature of $\approx 130$~\degree C, by means of polarized neutron reflectometry (PNR) in order to reveal its magnetic field-temperature phase diagram. In bulk this material has a helimagnetic ground state. Here we use PNR data to show that the helimagnetism survives in a CoZn film only 100~nm thick, roughly half a helical period. Given the subtlety of differentiating the helical state from a uniform state with reduced interfacial magnetism, we have developed and employed a Bayesian approach that uses model selection to analyze our PNR data and to map out the field-temperature phase diagram of our thin film.      
\end{abstract}

\maketitle


\section{Introduction}
\label{sec:intro}

\lettrine[lines=3]{T}{he} archetypal materials for the study of skyrmions are the magnetic materials with the B20 crystal structure \cite{Nagaosa2013}. This class includes 3d transition metal silicides and germanides, the foundational materials being the monosilicide MnSi \cite{Williams1966} and monogermanide FeGe \cite{Lundgren1968,Wappling1968}. Interest in these materials was spurred by the resolution of their unusual magnetic properties being due to a long range magnetic state identified as a helical magnetic state (`helical spin density wave') \cite{Lundgren1970, Ishikawa1976}. Notably, small angle neutron scattering (SANS) was used by Ishikawa \emph{et al.} \cite{Ishikawa1976} for MnSi to directly measure the helix wavelength. Later, the helical wavelength in FeGe would be measured, again using SANS \cite{Lebech1989}.

From here new ternary silicide compounds were explored, chiefly FeCoSi and CoMnSi \cite{Beille1983} which also showed long range helical magnetic order. It followed that topological `vortex' states were predicted in condensed matter systems \cite{Bogdanov1989, Bogdanov1994}, and in 2006 magnetic materials with chiral interactions were predicted to be capable of hosting skyrmions \cite{Rossler2006}. This was soon followed by the breaking of new ground with the publication by M\"{u}hlbauer \emph{et al.} \cite{Muhlbauer2009c}, which reported on their findings of skyrmions in a magnetic system, namely in the MnSi system measured once again using SANS. The confirmation of the predicted skyrmions in a magnetic metal lead to a surge in related research across the B20 class and beyond. Within bulk systems SANS is used extensively to directly measure magnetic helices \cite{Grigoriev2006,Grigoriev2007,Grigoriev2009,Grigoriev2014}. 

None of the materials described above exhibit robust magnetic order at or above room temperature. It has long been known that CoZnMn alloys, which are room temperature ferromagnets, possess a $\beta$-Mn phase \cite{Buschow1983,Hori2007,Karlsen2009,Xie2013c}. This is a chiral structure with space group $P4_332$ or $P4_132$, depending on the handedness. Around a decade ago, CoZnMn was shown, in the bulk, to host skyrmions at room temperature \cite{Tokunaga2015} and a flurry of other studies since that time have explored their properties \cite{Karube2016a,Karube2017b,Takagi2017,Morikawa2017a,Karube2018,Bocarsly2019}. These metastable skyrmion phases \cite{karube_metastable_2020}, as in the B20 materials, are based on a helimagnetic ground state. 

The origin of these helical magnetic structures is explained by the Dzyaloshinskii-Moriya interaction (DMI) which arises from the chirality of the crystal structure \cite{Nakanishi1980,Bak1980}. The interplay of Heisenberg exchange, which alone typically gives rise to collinear states, with the DMI gives rise to a chiral canting between neighboring spins and hence a long-range spin helix. The helimagnetic state is often described by a propagation wavevector \textbf{Q}, about which a full rotation of the spin is one wavelength of the helix:
\begin{equation}
Q=\frac{2\pi}{\lambda_\mathrm{h}},
\label{Eqn:HelixQ}
\end{equation}
where $\lambda_\mathrm{h}$ is the helix wavelength. This rotation, and thus wavelength, is defined by the ratio of the Heisenberg exchange energy, $J$, and the DMI constant, $D$, given by:
\begin{equation}
 \lambda_\mathrm{h} = 2\pi \left(\frac{J}{D} \right) ;  
\end{equation}
knowledge of the helix wavelength gives information about the relative strength of the two interactions in the helimagnetic system. 

Given the potential usefulness of skyrmions in spintronics \cite{Fert2017,Marrows2021}, it is desirable to prepare these materials in a thin film form. We have previously reported the growth of thin films of CoZn with a $\beta$-Mn crystal structure with a Curie point exceeding room temperature \cite{Deargthesis,Dearg2025}. Here we report the detection of helimagnetic ordering in these films, in spite of their polycrystalline nature, using the technique of polarized neutron reflectometry (PNR) \cite{zhubook}, which provides a magnetic depth profile of the film.

Within a thin-film it becomes necessary to refine this understanding of the helical spin structure. In the bulk, one would expect the helical axis to reorient itself relative to an applied field, however in a thin film we are constrained by the shape anisotropy. In this instance we adopt the term ``helicoid'' to describe the magnetic structure in a helimagnetic film as it extends through the plane(s) of the film \cite{Karhu2012,Porter2015,Wilson2014}. The helicoid describes the resulting structure of a helimagnet under an applied field, $H$, perpendicular to the helical axis, $Q$, while constrained by anisotropy, $K$. This causes the helix to distort, with moments canting towards the the applied field direction, instead of reorientating the axis, which effectively unwinds the helix thus changing the period. This behavior, where the helix wavevector Q is fixed normal to the film plane and the helix unwinds with applied in-plane field, has been seen to occur in thin-films of both MnSi \cite{Karhu2012,Wilson2013} and FeGe \cite{Porter2015}.

\begin{figure}[t]
\centering
    \begin{tikzpicture}
        \node at (0, 0){ \includegraphics[width=238pt]{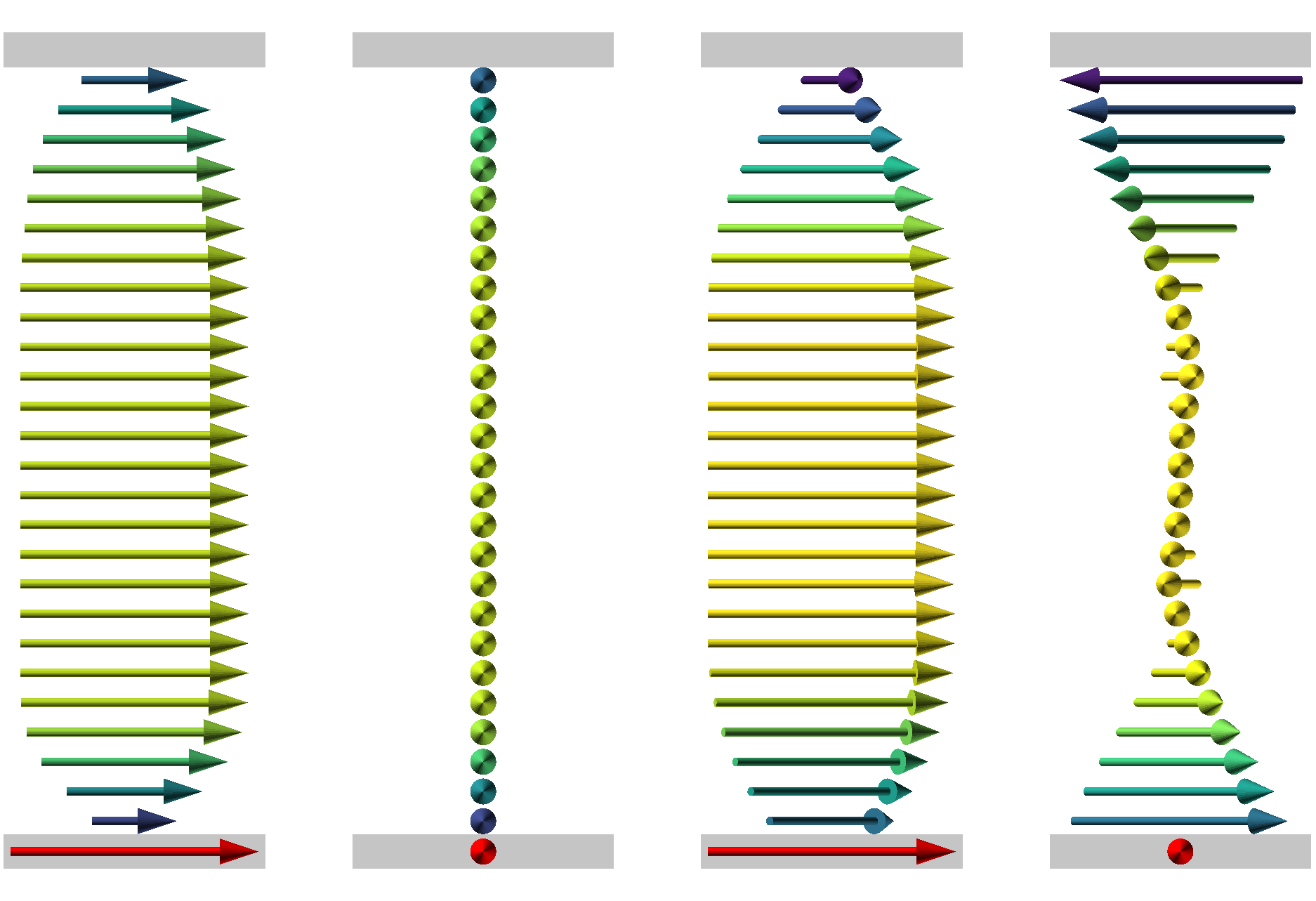}};
        \node at ( -3.94, 2.97){\bf{(a)}};
        \node at ( -1.71, 2.97){\bf{(b)}};
        \node at ( 0.484, 2.97){\bf{(c)}};
        \node at ( 2.723, 2.97){\bf{(d)}};
    \end{tikzpicture}
	\caption{
    Spin configurations for a helimagnetic thin film of CoZn , where the helix axis is constrained by shape anisotropy \cite{Wilson2013}. The grey planes indicated are the substrate-film interface (bottom) and the film-atmosphere interface (top), with magnetic field $H$ applied in the plane of the film (shown by the direction of the red arrow), which acts to distort the helix.
    (a) Illustrates a `uniform' model where there is a reduction in moment magnitude near the interfaces, and the depth dependent moment has no transverse moments.
    (b) On-axis to the applied field view of (a) demonstrating no transverse moments in the plane of the film.
    (c) Illustrates a `helix' model where the depth dependent moment is distorted into a helicoid the direction of the field.
    (d) Illustrates how the helical solution to the PNR signal would twist in-plane from substrate to surface, with the substrate and surface near-anti-aligned from each other. The profiles are based on the PNR results presented below in Fig.~\ref{fig:pnr_sld_25C_10mT} (25~\degree C temperature, 10~mT applied field).
    \label{fig:helicoid}}
\end{figure}

The film we have studied is approximately $1000$~\AA\ thick. This is roughly half of the bulk helical wavelength reported for Co$_{10}$Zn$_{10}$ \cite{Tokunaga2015}, and so it is not possible to fit even a full turn of the spin helix into the thickness of the film. Identification of helical order is therefore a significant experimental hurdle, since the most that can be expected is subtle deviations from a state that is otherwise uniformly magnetized along the field direction. The challenge is illustrated in Fig.~\ref{fig:helicoid}. The PNR technique is sensitive only to the component of magnetization along the field direction, unless the highly time-consuming polarization analysis variant is used. Panels (a) and (c) depict this component for two spin configurations: a `uniform' state in which the magnetization is reduced in magnitude somewhat, as often happens due to effects such as intermixing of oxidation; and the `helical' state that we wish to identify where the spins are distorted into a helicoid by the applied field. These profiles are very similar. Panels (b) and (d) show the other in-plane component of these two spin configurations. The presence of transverse moments in (d) confirms the presence of the helicoid. Here we describe the use of a Bayesian approach to the analysis of the PNR data that uses model selection to map out the magnetic field-temperature phase diagram of our film and show that the presence of helimagnetism is overwhelmingly probable below the saturation field and Curie point of our film. The helical wavelength $\lambda_\mathrm{h}$ is determined to be ${\sim}1700$~\AA\  from these measurements, only slightly shorter than the reported bulk value.

\section{Experimental Methods\label{sec:expt}}

\subsection{Sample Growth/Preparation\label{sec:growth}}

Thin film samples of amorphous/polycrystalline cobalt-zinc were deposited by co-sputtering from elemental targets. These are grown on a 1000~\AA\ SiO$_{2}$ surface layer of a 0.5~mm thick thermally oxidized Si substrate. This is done in balanced magnetrons with an argon working gas at a growth pressure of 3.2~mTorr. The deposition system has a base pressure of approximately $10^{-8}$~Torr. The deposition power of the Co and Zn guns were controlled separately to set the stoichiometry of the deposited alloy film.

To promote the growth of the $\beta$-Mn crystal structure, we annealed the sample post-deposition. The sample was isolated in a quartz glass ampule, repeatedly pump-purged with 99.9999~\% pure argon gas before being evacuated to a roughing pressure of order $10^{-1}{\sim}10^{-2}$~Torr.

The annealing profile staggered decreasing ramp rates of 5/3/1~\degree C per minute relative to temperatures of 300/350/400~\degree C, with relative dwell times of 2/1/72~hours; before finally being cooled at the natural rate of the furnace with the output turned off.

Full details of sample preparation methods are described in Refs \onlinecite{Deargthesis} and \onlinecite{Dearg2025}.

\subsection{X-Ray Reflectometry/Diffraction\label{sec:xrs}}

X-ray reflectometry (XRR) and diffraction (XRD) were performed using a Rigaku Smartlab rotating anode diffractometer. The diffractometer was set up in a Bragg-Brentano geometry with a Cu anode source  and Ge (220) 2-bounce monochromator, and was run in continuous scanning mode at 0.5~deg/min with a resolution of 0.0002~deg for both XRR and XRD.

\subsection{Magnetometry\label{sec:squid}}

Magnetometry was performed using a superconducting quantum interference device vibrating sample magnetometer (SQUID-VSM), equipped with an oven module option which operates the sample chamber at a pressure of ${<}100$~mTorr. This option allows for temperatures $T$ up to 700~\degree C to be safely achieved by way of direct contact of the heating element with the sample.

We measured a series of magnetization ($M$) hysteresis loops ($M(H)$) from a $\mu_0 H = +700$~mT saturating field to $-700$~mT and back again. The hysteresis loops were obtained at the temperatures of 26.85, 105.85, 126.85, and 155.85~\degree C, where the final loop was acquired well above the Curie temperature of CoZn of $T_\mathrm{C} \approx 129$~\degree C. These were close to the PNR measurement temperatures of 25, 106, 127, and 156~\degree C.

\subsection{Polarized Neutron Reflectivity\label{sec:pnr}}

The depth dependent magnetization profiles were measured using PNR on the POLREF beamline at the ISIS Neutron and Muon Source. Sample heating was conducted using a bespoke vacuum furnace (base pressure of ${\sim}10^{-7}$~mbar), which was itself mounted within an electromagnet using a 100~mm pole gap. The sample was mounted onto a copper mounting block with a thin layer of thermal conducting grease before being loaded into the furnace and pumped down to ${<}10^{-5}$~mbar prior to heating and measurement.

The sample was set into its initial magnetic state by heating to 156~\degree C (above the Curie temperature, $T_\mathrm{C} \approx 129$~\degree C) under a 700~mT applied field, then field cooled to the required temperature in an applied field of $\mu_0 H = 0.9$~mT. At this point PNR data were taken out to a maximum wavevector transfer of $Q_{z} = 0.12$~\AA$^{-1}$. Primary isotherms scans were taken at 25~\degree C and 106~\degree C, and a primary isofield scan at $\mu_0 H=0.9$~mT. For each isotherm, the field was increased incrementally up to a maximum value of $\mu_0 H = 700$~mT, a value known from hysteresis loop measurements to be sufficient to magnetically saturate the sample \cite{Dearg2025}. The system was reset as required by heating above the Curie temperature and field cooling to the next required value in an appropriately low guide field to not depolarize the neutron beam. This allowed monitoring with PNR to determine when the system had reached thermal equilibrium and a stable neutronic sample alignment was obtainable. 

\section{Results\label{sec:results}}

\subsection{X-Ray Reflectometry/Diffraction\label{sec:xrrxrd}}

We studied a CoZn film deposited using sputter powers of 36~W for Co and 25~W Zn, close to the optimal ratio to obtain the $\beta$-Mn structure without excess Co particulates \cite{Dearg2025}, with nominal thickness of 1000~\AA. As the helical period is reported to be 1850\textendash2100~\AA\ for Co$_{10}$Zn$_{10}$ $\beta$-Mn in bulk \cite{Tokunaga2015}, this thickness then represents approximately half of the bulk period. 

\begin{figure*}[t]
\centering
    \begin{tikzpicture}
        \node at (0, 0){ \includegraphics[width=\textwidth]{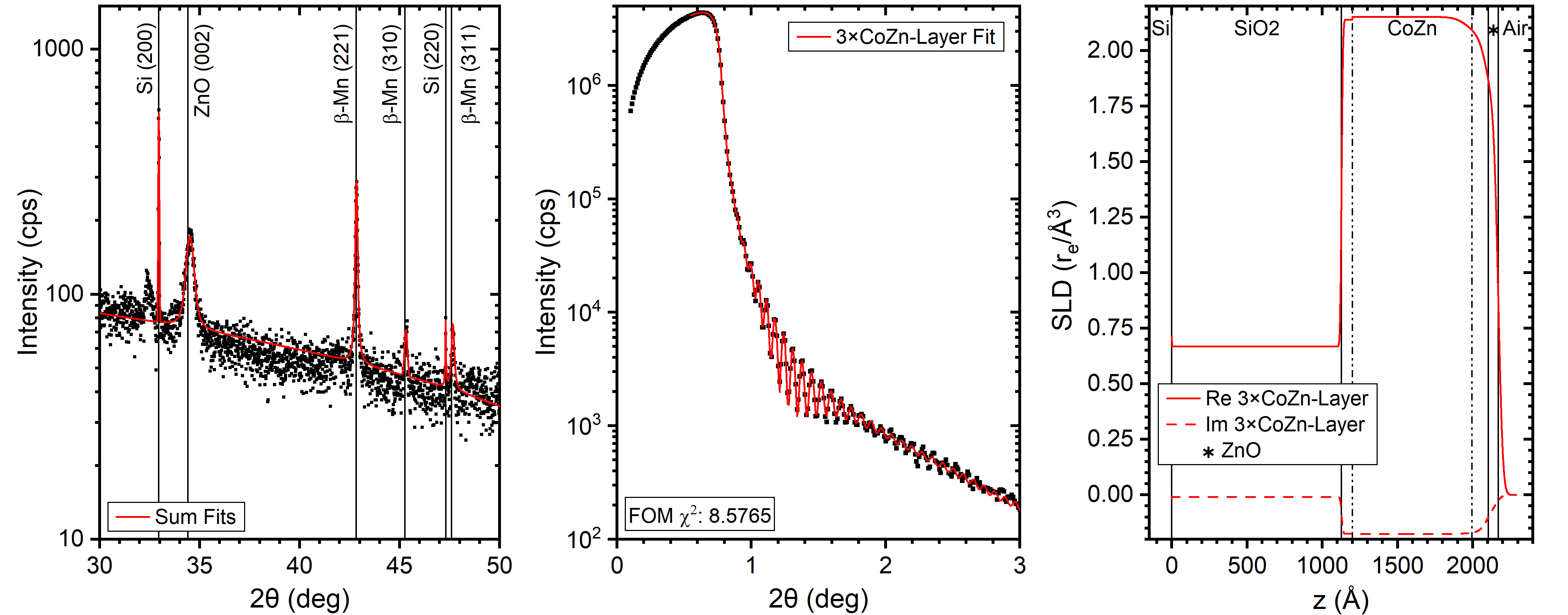}};
        \node at (-8.56, 3.35){\bf{(a)}};
        \node at (-2.73, 3.35){\bf{(b)}};
        \node at ( 3.25, 3.35){\bf{(c)}};
    \end{tikzpicture}
    \caption{X-ray analysis of the annealed film of CoZn, measured using Cu $K_\alpha$ radiation (wavelength 1.541~\AA).
    (a) XRD pattern showing diffraction peaks overlaid with the sum of individual peak fits using a Voigt profile and a linear background. Relevant peaks appear at $\sim$34.4\degree\ ($\langle 002 \rangle$ ZnO) and $\sim$42.7\degree\ ($\langle 221 \rangle$ $\beta$-Mn), with some hints of higher order peaks at $\sim$45.2\degree\ and $\sim$47.4\degree\ ($\langle 310 \rangle$ \& $\langle 311 \rangle$ $\beta$-Mn, respectively). The peak at $\sim$32.9\degree\ is the classically ``forbidden'' $\langle 200 \rangle$ peak of Si. 
    (b) XRR results for the film with fit as described in the main text. 
    (c) Comparison of XRR determined SLDs for the 3-layer model.
    \label{fig:xrs}}
\end{figure*}

The XRD pattern of the post-annealed film, presented in Fig.~\ref{fig:xrs}(a), confirms the expected predominant $\langle 221 \rangle$ $\beta$-Mn texture alongside the presence of the ZnO capping layer \cite{Dearg2025}. 

Fig.~\ref{fig:xrs}(b) displays the corresponding XRR measurement, featuring clear and regular Kiessig fringes indicative of high structural smoothness. The overall fringe intensity envelope is consistent with the presence of a capping layer. Fitting of the XRR data was performed using the \textsc{GenX} code \cite{Bjorck2007}, utilizing a model in which the active CoZn layer is split into three sub-layers: an interfacial layer adjacent to the substrate, a bulk layer, and a top layer interfacing with the ZnO cap.

The resulting fit yields a total CoZn film thickness of ${\sim}976$~\AA\ and an interfacial roughness of $\sim44$~\AA. The analysis indicates a minimal surface ZnO thickness (${\sim}66$~\AA), which is comparable to the surface roughness of the underlying CoZn film (${\sim}44$~\AA). The scattering length density (SLD) depth profile extracted from the fit is shown in Fig.~\ref{fig:xrs}(c). This reveals a slight SLD depletion in the CoZn film at the interface with the SiO$_2$ substrate (${\sim}0.55$\% reduction extending across ${\sim}72$~\AA), but with the vast majority of the film consisting of the bulk behavior.

Overall, these structural results confirm that the film possesses the intended morphology of a $\beta$-Mn ordered layer of CoZn with a ZnO capping layer \cite{Dearg2025}, with a high degree of interfacial smoothness that renders it suitable for subsequent PNR analysis. 

\subsection{Magnetometry\label{sec:mvhmvt}}

\begin{figure}[t]
\centering
    \begin{tikzpicture}
        \node at (0, 0){ \includegraphics[width=\columnwidth]{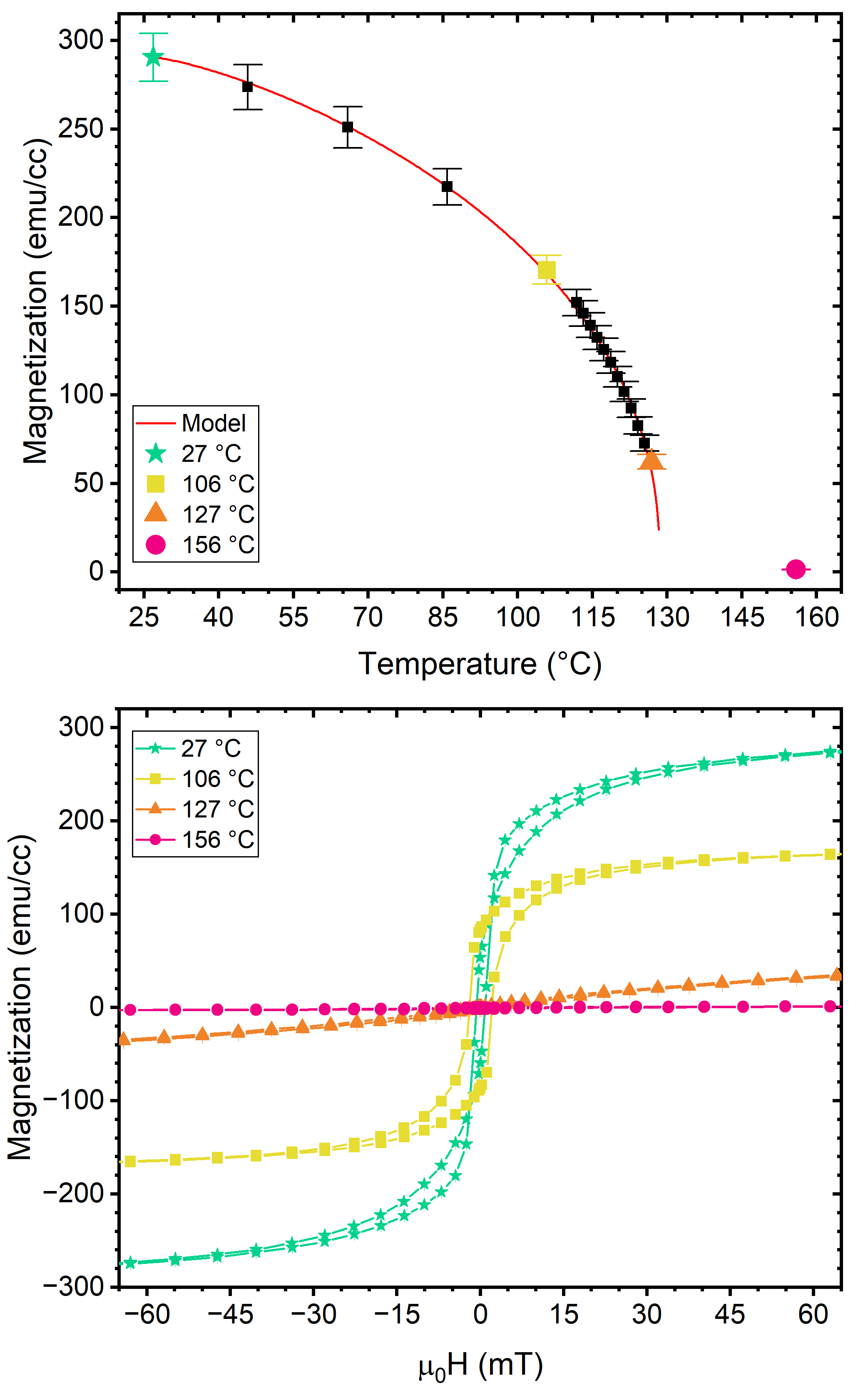}};
        \node at (-4.11, 6.55){\bf{(a)}};
        \node at (-4.11, -0.36){\bf{(b)}};
    \end{tikzpicture}
    \caption{SQUID-VSM characterization results. 
    (a) $M(T)$ shows the expected form with a $T_\mathrm{C}$ of ${\sim}128.6^{+0.3}_{-0.1}$~\degree C. 
    (b) Isothermal $M(H)$ hysteresis loops at temperatures close to those subsequently used for PNR.
    \label{fig:squid}}
\end{figure}

The SQUID-VSM results for the film are summarized in Fig.~\ref{fig:squid}. Panel (a) shows the magnetization vs. temperature ($M(T)$) curve, confirming the usual magnetic properties of $\beta$-Mn CoZn with a $T_\mathrm{C} = 128.6^{+0.3}_{-0.1} $~\degree C and no retained magnetism above that temperature \cite{Dearg2025}. This was determined by fitting with the empirical interpolation of Bloch's law given in Ref.~\onlinecite{Evans2015}, modified to limit the fit to the above room-temperature ($T_{300} \to{} T_{C}$) region, given by:
\begin{equation}
M(T)=M_{300}\times\left(1-\left(\frac{T}{T_{\text{C}}}\right)^{\alpha}\right)^{\beta},
\label{eqn:CurieBloch}
\end{equation}
where $M_{300}$ is the magnetization at room temperature, the $\alpha$ term captures the behavior around the `low' temperature region, and the $\beta$ term captures the behavior near the Curie point ($T_C$). For a protoypical ferromagnet, $\alpha = \frac{3}{2}$ and $\beta = \frac{1}{2}$.

Fig.~\ref{fig:squid}(b) shows isothermal $M(H)$ hysteresis loops at temperatures following those of the PNR that we shall discus below, with the field applied in-plane. We observed that the sample is always saturated at a field of $\mu_0 H = 700$~mT, and so we expect that a PNR measurement at that field should be consistent with a uniformly magnetized state at any temperature ${<}T_\mathrm{C}$. The field values at which we see a reduction in $M$ below the saturated value are those at which we should make PNR measurements seeking evidence for helimagnetism.

\subsection{Polarized Neutron Reflectivity\label{sec:Polarized_neutron_reflectivity}} 

PNR was used to investigate the depth dependent in-plane magnetization profile with respect to temperature and field. Due to its magnetic depth sensitivity, PNR is an ideal probe to investigate the depth dependent in-plane magnetic texture present within the sample. Here, PNR was used to determine if the sample was in one of two magnetic states - helical or uniform, as illustrated in Fig.~\ref{fig:helicoid} - at various points in phase space. We measured our sample at various values of field whilst at fixed temperatures of 25, 46, 66, 86, 106, and 156~\degree C, yielding a series of data sets along isotherms.

Depth dependent structural and magnetic information is obtained from the PNR data by fitting parametric models that describe the SLD, which has both structural (nuclear) and magnetic  contributions. Since the structure was assumed to be invariant in the temperature range investigated in this study, the structural SLD was determined from a large co-fit of the most of the data taken for this study (details in the appendix section \ref{sec:Appendix_PNR_struc_fit}). Subsequently, the structure was fixed while fitting the uniform and helical models to each data set - by doing this, the goodness of fit could be purely assessed on changes in the magnetic scatting length density (mSLD) profile.

The helical magnetic profile used in this study is that of a distorted helicoid magnetization $M(z)$, similar to what is described in Refs \onlinecite{Porter2015} and \onlinecite{rossler2011chiral}:
\begin{equation}
    M(z) = A_{1}\sin\left(\frac{2\pi z}{\lambda_\mathrm{h}} + \phi \right) + A_{2}\cos^{2}\left(\frac{2\pi z}{\lambda_\mathrm{h}}+\phi\right).
    \label{eqn:distorted_helicoid}
\end{equation}
The representation given by Eq.~\ref{eqn:distorted_helicoid} describes the projection of the magnetization along the applied field direction as a function of depth $z$, where $\rho_{\mathrm{M}}$ is the mSLD, $A_1$ and $A_2$ are the amplitudes of the undistorted helix and the component representing the distortion of the helix (as a result of an elliptical component in the $x-y$ plane), respectively. $\lambda_\mathrm{h}$ is the wavelength of the undistorted helix (i.e. when $A_2 = 0$) and $\phi$ is the phase offset of the profile with respect to depth in the CoZn layer.

The uniform magnetic profile was modeled with a region of non-varying mSLD across the CoZn layer. The bottom interface was modeled using the standard error function interface profile - the value of which was made independent from the structure. The top interface was modeled with a parametric spline with 2 control points, the positions of which were fixed to the structural values for that layer (details of this aspect of the modeling are described in the appendix section \ref{sec:Appendix_unifrom_mag_profile}).

\begin{figure*}[tbh]
    \includegraphics[width=\textwidth]{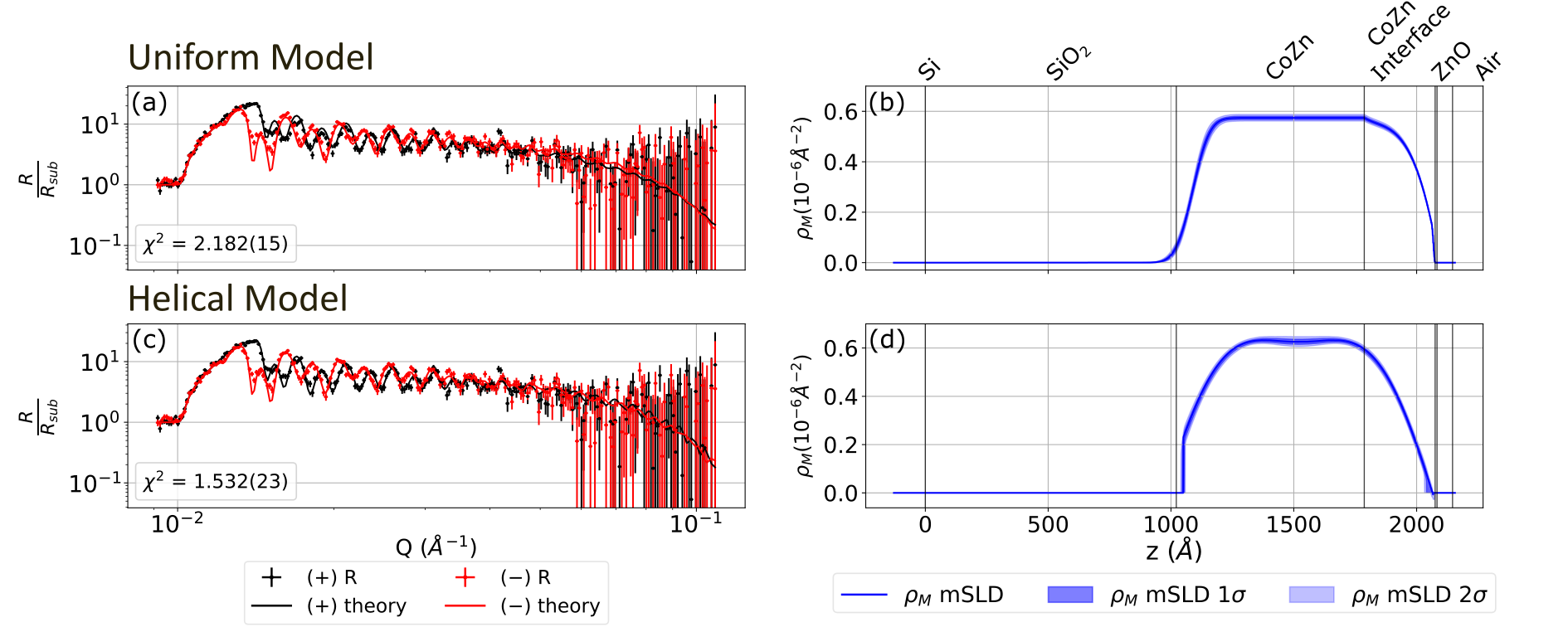}
    \caption{PNR data acquired at 25~\degree C and an applied field of 10~mT, normalized by the substrate reflectivity and fits for both uniform and helical magnetic models. 
    (a) PNR data and fit for uniform magnetic model. 
    (b) Magnetic SLD profile resulting from the fit in (a). 
    (c) PNR data and fit for helical magnetic model. 
    (d) Magnetic SLD profile resulting from the fit in (c). 
    The shaded areas provide the 1 and $2\sigma$ confidence intervals for the profile.
    \label{fig:pnr_sld_25C_10mT}}
\end{figure*}

Fig.~\ref{fig:pnr_sld_25C_10mT} shows a comparison between the uniform (panels (a) and (b)) and helical (panels (c) and (d)) magnetization profiles described above for the 25~\degree C, 10~mT data set. Panels (a) and (c) show the PNR data and best fit lines, while (b) and (d) show the depth dependent magnetization (mSLD) profiles derived from the fits. At first glance, the two mSLD profiles have broadly similar maximum magnitudes and extent in depth - albeit with the helical model exhibiting a higher maximum mSLD than the uniform model. This is somewhat expected since both models capture the maximum splitting between the spin up (black) and spin down (red) reflectivity curves at the critical edge (this feature in the data is strongly dependent on the total magnetization in the sample). Observing the fits to the data more closely, the uniform model fails to capture the damping in the fringe minima in the same way that the helical model does. The helical model mSLD profile inherently has less sharp changes in mSLD with respect to depth, which would manifest as an effective greater magnetic roughness as probed by the neutrons, resulting in a more rapid damping of the Kiessig fringes, than a sharper magnetic interface. Consequently, the helical magnetization profile provides a significantly better fit to the data - $\chi^{2} = 1.532(23)$ as opposed to $\chi^{2} = 2.182(15)$ for the uniform magnetization profile, indicating that for this field and temperature point, the helical model is the more probable interpretation of the data. Moreover, from the $\rho_{\text{M}}$ data presented in Fig.~\ref{fig:pnr_sld_25C_10mT}(d) we can calculate a layer averaged magnetization of $\approx 170$~emu/cc. This value is in excellent agreement with the 25~\degree C layer averaged magnetization as measured by SQUID in Fig.~\ref{fig:squid}(b). 

Closer observation of the helical profile shows that the helical profile is not centered within the layer - indeed a phase shift of $-0.19\pm0.05$~rad was yielded from the fit. The negative value of the phase indicates that the profile is shifted towards the bottom interface. This could point towards the profile being anchored to the $\mathrm{SiO_2/CoZn}$ interface, possibly linked to the higher structural sharpness of this interface.  

The same comparison as shown in Fig.~\ref{fig:pnr_sld_25C_10mT} was conducted for all data taken within this study: both helical and uniform models were fitted to the data for each value of field and temperature at which the sample was measured. A common method of performing model selection is to compare the goodness-of-fit (in our case reduced $\chi^{2}$) of the best fit parameters. However, there are issues with purely using a goodness-of-fit metric as a model selection criterion. For example, this methodology does not take into account variation in the model parameters allowed by variance in the data (error in reflectivity), which can in principle produce similar goodness-of-fit for quite different parameter sets. This is a problem that is particularly acute in reflectometry as a result of the inverse phase problem \cite{sivia2011elementaryChapter4}: the loss of the phase information allows for multiple solutions for the same reflectivity profile, i.e. a loss of uniqueness. Here we use the Bayesian model selection approach of calculating and comparing the Bayesian model evidence \cite{knuth2015bayesian, farrell2018chapter11, trotta2008bayes, hicks2018bayesian} to indicate whether the helical or uniform magnetization profiles are more probable for each of the field-temperature conditions measured.

\subsection{Bayesian analysis of PNR data\label{sec:Bayes_PNR}}

\subsubsection{Model selection}

Bayesian model selection relies on comparing the probability of one model to another with respect to the data. In order to do this, we need to calculate the probability of each model to be compared ($M_i$) given the data ($d$) is true - otherwise known as the posterior probability $P(M_i|d)$ \cite{sivia2006data}. To compare 2 models, the ratio of the posterior probabilities - the posterior odds ratio - is calculated. For models $M_1$ and $M_2$, the posterior odds ratio is defined as \cite{knuth2015bayesian, hicks2018bayesian}:
\begin{equation}
        \frac{P\left(M_1|d\right)}{P\left(M_2|d\right)}=\frac{P\left(M_1\right)}{P\left(M_2\right)}\frac{P\left(d|M_1\right)}{P\left(d|M_2\right)},
        \label{eqn:posterior_odds}
\end{equation}
where $\frac{P\left(M_1\right)}{P\left(M_2\right)}$ are the prior odds and $\frac{P\left(d|M_1\right)}{P\left(d|M_2\right)}$ is the ratio of the model evidence for $M_1$ and $M_2$; also known as the Bayes factor $\cal{B}$. The value obtained from calculating the posterior odds ratio for two models $M_1$ and $M_2$, provides a metric whereby a strength of model preference can be quantified - numbers greater than 1 favor the model on the numerator, while numbers less than 1 indicate the inverse. In the case of this study, we have no prior preference for the helical or uniform magnetization models for any given measurement condition, so the prior odds are assumed to be 1. The posterior odds ratio is then just simply the Bayes factor $\cal{B}$. Bayes factors were calculated for all measurement conditions taken in this study and used to determine a model preference of helical or uniform in each case.

The strength of using Bayes factors to inform model selection preferences lies in the fact that it is the ratio of the model evidence for the pair of models being compared. The model evidence (also known as the marginal likelihood) $P(d|M)$, can be thought of as the probability of seeing the data $d$ in the entire parameter space defined in model $M$ \cite{farrell2018chapter11}. It is calculated by integrating the product of the likelihood $P(d|\theta, M)$ and the prior $P(\theta|M)$ across all parameter space:
\begin{equation}
    P(d|M) = \int P(d|\theta, M)P(\theta|M)d\theta,
    \label{eqn:model_evidence}
\end{equation}
where $\theta$ is a set of parameter values in the model $M$. As a result, the model evidence is the prior weighted average of all fits in parameter in model $M$. The evidence calculation also encodes an Occam's factor (see Ref. \onlinecite{knuth2015bayesian}), that takes into account model complexity; models with greater flexibility and/or larger prior volume require a larger maximum likelihood to offset the increased model complexity. In the present context, the helical model is more flexible with respect to the range of mSLD profiles and shapes that can be generated in comparison to the uniform model - for the helical model to be preferred it must provide a significantly better fit to the data in comparison to the uniform model. 

The values for the model evidence obtained in this study, were calculated using the UltraNest nesting sampling Python package \cite{buchner2016statistical, buchner2019collaborative, Buchner2021} in a bespoke setup with the Refl1D reflectometry fitting software package \cite{refl1d1p0p0_2025, bumps1p0p3_2025}. As indicated above (Eq.~\ref{eqn:model_evidence}), parameter priors can have a significant influence on the evidence calculation. In order make the evidence calculation as reliable as possible, all shared model priors (parameters that were used in both models) were identical and used identical fit ranges. In this way, the models were only defined by the parameters that make them unique from each other, thereby minimizing the effect of inconsistent priors on the evidence calculation.

\begin{figure}
    \includegraphics[width=8cm]{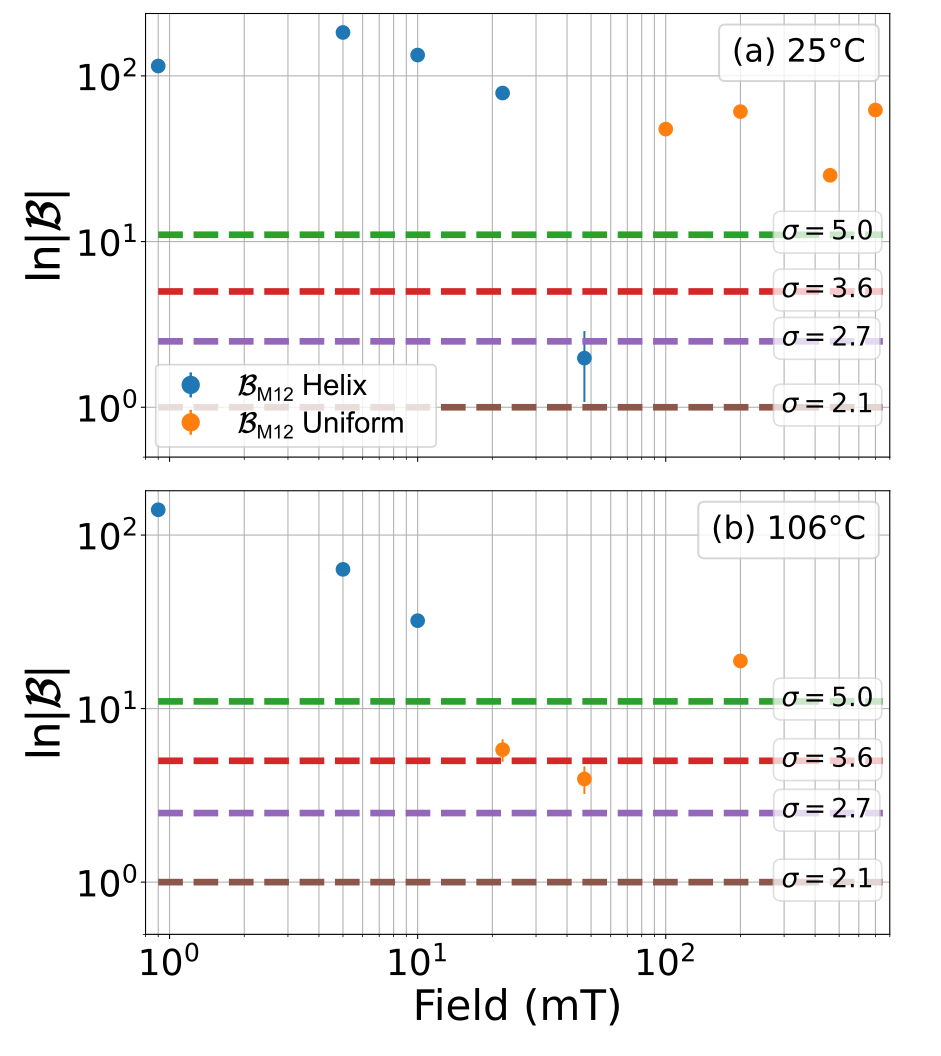}
    \caption{Bayes factors for both uniform and helical models for each measured field value at (a) 25~\degree C and (b) 106~\degree C isotherms. These are taken as the ratio between helical ($M_\mathrm{helix}$) and uniform model ($M_\mathrm{uniform}$) evidence values. Both ratios ($M_\mathrm{helix}/M_\mathrm{uniform}$ and $M_\mathrm{uniform}/M_\mathrm{helix}$) were calculated, however, only Bayes factors greater than 1 are shown here - i.e., showing the more probable model (Bayes factors of less than 1 were omitted for clarity). Model priors were assumed to be equal for both models. Thresholds for $\sigma = 5.0$, 3.6, 2.7, and 1.0 are indicated by the dashed lines.
    \label{fig:Bayes_factors_25C_106C}}
\end{figure}

Fig.~\ref{fig:Bayes_factors_25C_106C} shows the calculated Bayes factors ${\cal B}_{M12}$ and ${\cal B}_{M21}$ for the 25~\degree C and 106~\degree C isotherms, defined as:
\begin{eqnarray}
    {\cal B}_{M12} &=& \frac{P\left(d|M_\mathrm{helix}\right)}{P\left(d|M_\mathrm{uniform}\right)},~\text{and} \\
    {\cal B}_{M21} &=& \frac{P\left(d|M_\mathrm{uniform}\right)}{P\left(d|M_\mathrm{helix}\right)},
\end{eqnarray}
in which the two models that are compared are the helical state $M_\mathrm{helix}$ and the uniform state $M_\mathrm{uniform}$. Here we are only displaying Bayes factors of above 1, so that the model displayed in the figure for a given measurement condition is the more probable model; i.e. ${\cal B}_{M12}$ (blue dots) and ${\cal B}_{M21}$ (orange dots) for helical and uniform respectively. The horizontal lines indicate confidence $\sigma$ values. These have been included as a more familiar representation of probability and ease of interpretation (the $\sigma$ scales were obtained from Ref.~\onlinecite{trotta2008bayes}).

Fig.~\ref{fig:Bayes_factors_25C_106C}(a) shows the Bayes factors for the 25~\degree C isotherm, a temperature well below $T_\mathrm{C}$. For low fields up to and including 47~mT the helical model (blue dots) is the more probable model, while above 47~mT the uniform magnetization model (orange dots) becomes more probable. This observation suggests that at 25~\degree C, the phase boundary between the helical and uniform states exists between an applied field of 47~mT and 100~mT. All measurement conditions in the 25~\degree C isotherm are all in excess of $\sigma=5.0$, with the exception of the 47~mT field point. While at 47~mT the helical model is more probable than the uniform model, it is less probable than all other Bayes factors calculated in this isotherm - landing between $\sigma=2.1$ and $\sigma=2.7$. It might be tempting to suggest that this decrease in probability indicates that the phase boundary is closer to 47~mT than 100~mT. However, caution should be taken when interpreting Bayes factors since the calculation is dependent on the average likelihood (goodness-of-fit) obtained from each model that are being compared \cite{sivia2006data,knuth2015bayesian,farrell2018chapter11} -- i.e. there may be multiple reasons as to why for this field the average likelihood of the helical model is less here.

Fig.~\ref{fig:Bayes_factors_25C_106C}(b) shows the Bayes factors with respect to the applied field for the 106~\degree C isotherm, not far below $T_\mathrm{C}$. At this temperature, the helical model is more probable up to 10~mT, beyond which the uniform model is more probable - indicating that the phase boundary at this temperature lies between 10~mT and 22~mT. 

\begin{figure*}[tbph]
    \includegraphics[width=12.5cm]{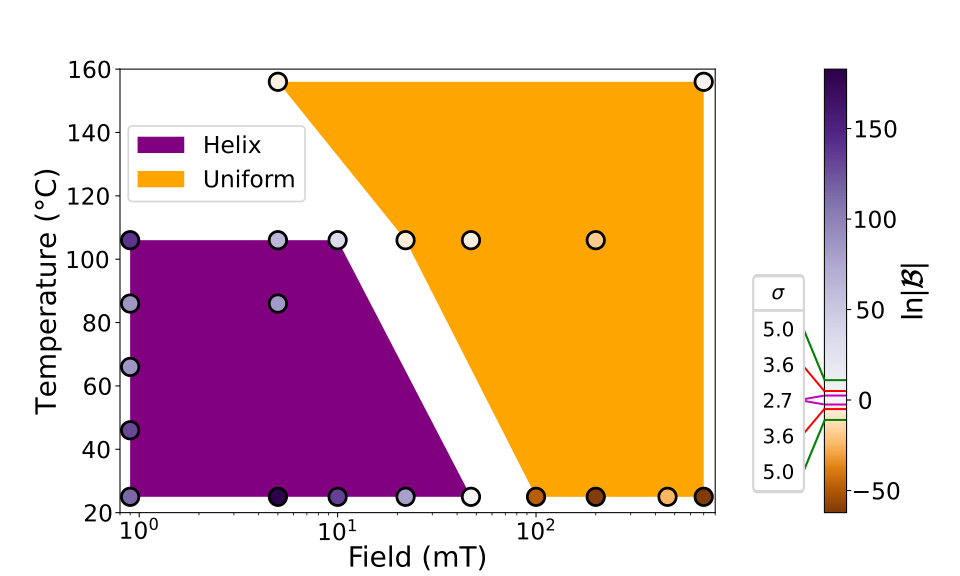}
    \caption{Phase diagram showing the helical and uniform magnetism regions derived from $\ln| \cal B|$. The color in the markers indicates how probable the particular data point is to be either helical or uniform - the deeper the color the more probable.
    \label{fig:Phase_diagram}}
\end{figure*}

By taking all of the field/temperature measurement points and calculating the Bayes factors comparing the helical and uniform models, a simple phase diagram can be constructed. Fig.~\ref{fig:Phase_diagram} gives this bi-modal phase diagram, showing the regions in phase space where the helical or uniform magnetization phases are more probable (purple and orange shaded regions respectively), where the phase contours are defined by the Bayes factor ($\ln|\mathrm{B}|$) of the data points that create the border of the contour. For ease of comparison $\ln|{\cal{B}_\mathrm{M12}}|$ only is shown on a continuous log scale - where positive $\ln|{\cal{B}_\mathrm{M12}}|$ indicates the helical model is more probable, while negative $\ln|{\cal{B}_\mathrm{M12}}|$ indicates the uniform model is more probable. The value of $\ln|{\cal{B}_\mathrm{M12}}|$ is encoded in the color of the data points - the deeper the color the more probable that data point is for the given model the color represents. 

The phase boundary has been deliberately left uncolored since, as discussed above, inferring the exact position from the Bayes factors alone is potentially misleading. The projected contours suggest that as temperature is increased the region in which the helical phase exists moves to smaller fields. Conversely, the uniform phase dominates at higher temperatures. In particular the highest measurement temperature of 156~\degree C, the film is above $T_\mathrm{C}$ and so the presence of helimagnetism is impossible, consistent with our findings.

\subsubsection{Helix pitch determination\label{sec:helix_pitch}}

\begin{figure*}[tbh]
    \includegraphics[width=15cm]{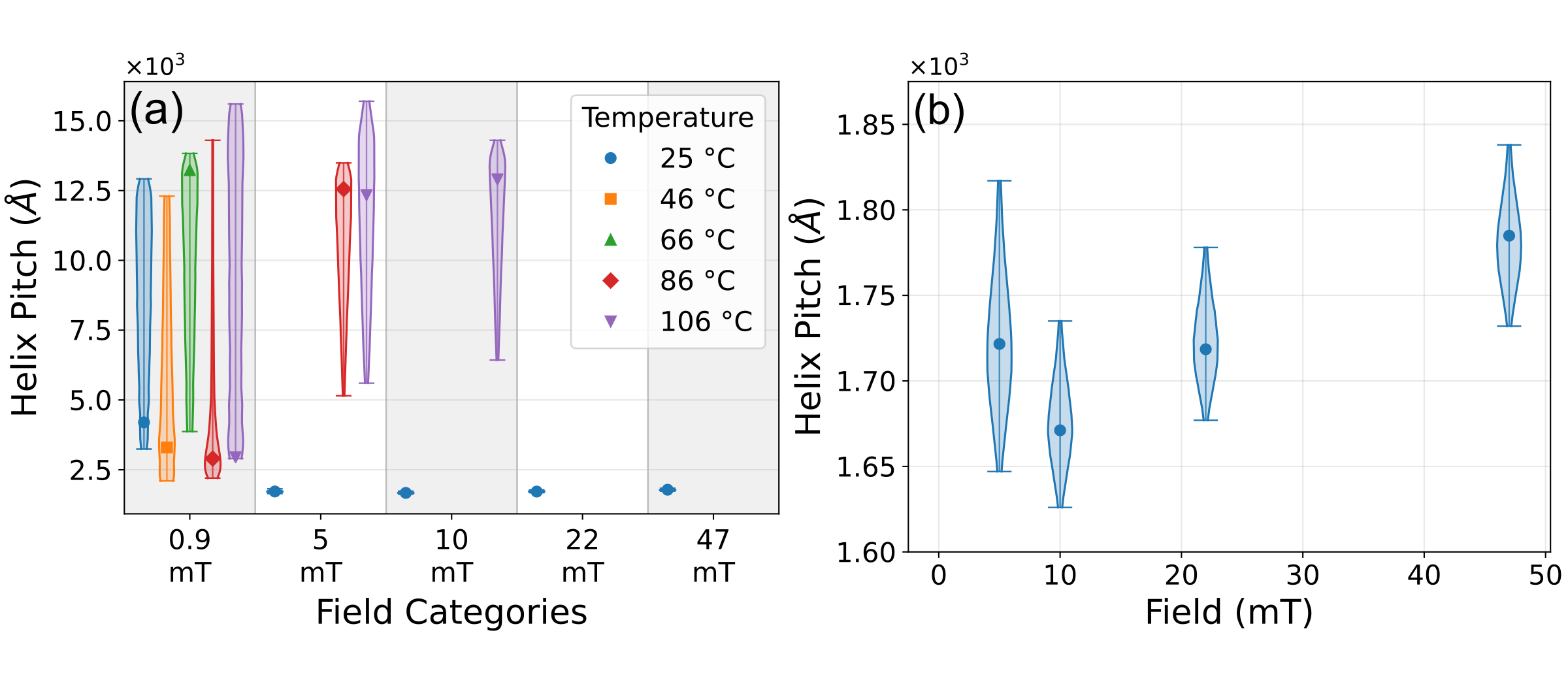}
    \caption{Helix pitch measurements. 
    (a) Category violin plot showing best fit values of helix pitch $\lambda_\mathrm{h}$ for measurement conditions where the helical model was determined to be more probable. The total range of the error bars are the 95\% credible interval of the total samples taken during the uncertainty analysis. The shaded width of the error bars represent the KDE of the posterior distributions obtained for the helix pitch for each of the measurement conditions. 
    (b) Scatter violin plot showing the best fit values of helix pitch $\lambda_\mathrm{h}$ for the 25~\degree C isotherm in the 5-47~mT applied field range. As in (a), the error bars are the 95\% credible interval of the samples taken. Here, the KDEs for all of the measurement conditions are broadly Gaussian, and the 95\% credible interval is over a smaller range than for the rest of the data shown in (a).
    \label{fig:helix_pitch_violin_plot}}
\end{figure*}

The process described in the previous subsection returns best fit values for the various physical parameters that described the sample. Fig.~\ref{fig:helix_pitch_violin_plot}(a) shows the best fit values of helix pitch $\lambda_\mathrm{h}$ for measurement conditions where the helical model was determined to be more probable. For most of the conditions, the 95\% credible interval is large, showing a high degree of uncertainty in the obtained value. This large degree of uncertainty is potentially a result of the small induced magnetization from the applied field. A small magnetization results in a small splitting between the spin up and spin down cross-sections in the PNR profile. Since the difference in the two spin states ultimately encodes the magnetization profile, certainty in that profile will be determined by how well these cross-sections are resolved. For a smaller splitting, a significantly longer count time would be required to decrease the uncertainty in determining the magnetic profile. The size of the 95\% credible interval reflects this uncertainty - where the best fit is falling in some cases at the edge of this range. 

In order to provide more information about the uncertainty shown, the kernel density estimation (KDE) of the posterior from which the uncertainty is derived, is also displayed. For some of the points (e.g. 106~\degree C, 0.9~mT), the uniform shape of the KDE shows the limited sensitivity that the helix parameter has on the likelihood of the fit. Conversely, the 86~\degree C, 0.9~mT data set shows the greatest degree of posterior weight around the best fit - even if the 95\% interval extends over a far larger range (i.e. has a long tail). For the most part the best fit values do land in the region of largest posterior weight, with the exception of 25~\degree C, 0.9~mT.

With the exception of the 25~\degree C, 0.9~mT dataset, the 25~\degree C isotherm demonstrated larger total magnetization values with the helical state persisting to higher applied fields. Fig.~\ref{fig:helix_pitch_violin_plot}(b) shows the best fit values obtained for the 25~\degree C isotherm in the 5-47~mT applied field range. The 95\% credible interval is significantly smaller for these measurement conditions than observed for the other helical datasets. Furthermore, the KDEs shown here are more Gaussian in shape, indicating a unimodal distribution. This means that the obtained parameter values are well defined. The improved certainty in these measurements allow a tighter determination of the helix pitch. Within the 95\% credible interval the helix pitch appears to be invariant within the 5-47~mT field range. Since the value of $\lambda_\mathrm{h}$ is set by the ratio of the Heisenberg exchange and Dzyaloshinskii-Moriya interactions, and will be unaffected by these modest fields for a given isotherm, this invariance gives us confidence that our Bayesian approach provides a realistic picture of the physics of the sample. 

\subsubsection{Helicoid profile}\label{sec:a1_a2_ratio}

\begin{figure}[tb]
    \includegraphics[width=8cm]{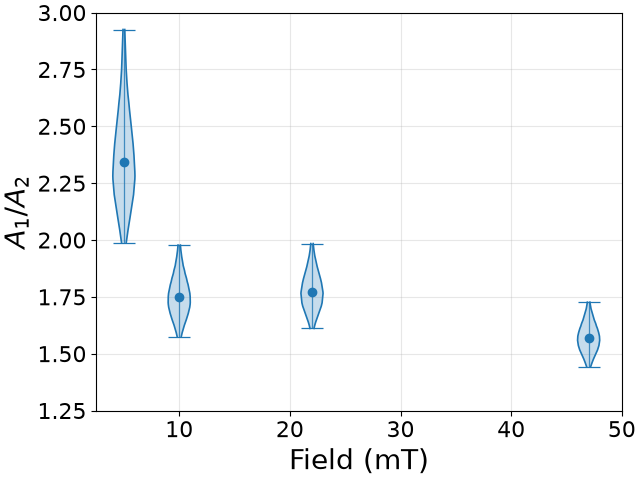}
    \caption{Scatter violin plot showing the ratio of the two amplitudes $A_1$ and $A_2$ of the helicoid for the 25~\degree C isotherm in the 5-47~mT applied field range. The error bars are the 95\% credible interval of the samples taken.}
    \label{fig:helix_a1_a2_ratio_violin_plot}
\end{figure}

The ratio between the two amplitudes $A_1$ and $A_2$ in Eq.~\ref{eqn:distorted_helicoid} provides information about whether the magnetization depth profile is a purely helical one (magnetic moment is sinusoidal with respect to depth), or a helicoid: the former is observed if $A_1$ is large and $A_2$ is negligible, the latter becomes a better description of the profile as $A_2$ increases. Fig.~\ref{fig:helix_a1_a2_ratio_violin_plot} shows the best fit values for the 25~\degree C isotherm for applied field values in the range of 5-47~mT. Observing the data, there appears to be a slight reduction in the ratio $A_1/A_2$ with increasing applied field. An increase of $A_2$ alters the profile such that for each rotation of the helix, the moment points more in the direction of the applied field as the strength of the field is increased. This would be consistent with the picture that at greater applied fields the helix transitions to a helicoid, which at high enough applied fields unwinds to a uniform magnetization profile.

As discussed in \S\ref{sec:helix_pitch}, the other field-temperature points where the helix was determined to be the more probable model displayed large uncertainties for $A_1$ and $A_2$ such that no meaningful trend could be observed for those isotherms. 

\section{Conclusion\label{sec:conc}}

To conclude, our main result is that we have observed helical ordering in this CoZn thin film by means of PNR coupled with a novel Bayesian approach to data analysis. The transition between ferromagnetic and the helical state is consistent with the bulk material, although the $T_\mathrm{C}$ appears to be lower (130~\degree C, consistent with Ref.~\onlinecite{Dearg2025}) than in the bulk (200~\degree C) \cite{Tokunaga2015}. The helix pitch $\lambda_\mathrm{h}$ is estimated to be $1724^{+68}_{-54}$~\AA\ (this is the average of the helix pitch values obtained at fields including and between 5-47~mT, where the error is defined by the average the upper and lower bounds of the 95\% credible intervals), and is invariant with respect to temperature and field.

Since the film thickness is comparable to half of the helix pitch, the departure of the $M(z)$ profile from that of a uniformly magnetized film is somewhat subtle. We have demonstrated the utility of Bayesian model selection in polarized neutron reflectometry by implementing this method to show that the helical state is overwhelmingly the most probable one for our film for fields below the saturation field at all temperatures below $T_\mathrm{C}$, allowing us to reject the model of uniform magnetization under these conditions. Comparison of the $A_1/A_2$ ratio indicates potential trend of increasing component of distorted helix with increasing field, which is consistent with observations in FeGe reported in Ref.~\onlinecite{Porter2015}. This makes sense when thinking about the applied field pulling moments into the field direction - slowly shifting from a helix to a helicoid - which eventually becomes a uniform magnetization state.

Now that we have established that sputtered thin films of $\beta$-phase CoZn are capable of showing helical ordering, we hope that our work will promote real-space imaging of chiral spin textures such as skyrmions--e.g. by Lorentz microscopy--as a stepping stone to room temperature planar devices based on spin textures such as skyrmions in this fascinating material.

\begin{acknowledgments}
This work was supported in part by the ISIS Facility Development Studentship scheme. We thank the ISIS neutron and muon facility for the provision of beamtime: RB1920567, and the use of the R53 Materials Characterisation Laboratory. Computing resources were provided by STFC Scientific Computing Department’s SCARF cluster. A. J. Caruana would like to thank N. Leo for helpful discussions.
\end{acknowledgments}

\section*{\label{sec:DataAvailibility}Data Availability}

The data associated with the structural and magnetic characterization of the sample described in this paper are openly available at the Research Data Leeds repository~\cite{dataleeds}. The PNR data are openly available from ISIS~\cite{dataisis}.

\section*{Author Contributions}
M. Dearg and A. J. Caruana contributed equally to this work.
M. Dearg - Manuscript original draft (lead), experimental planning, data acquisition and analysis (XRR, magnetometry, XRD, PNR).
A. J. Caruana - Manuscript original draft (lead), experimental planning, data acquisition and analysis (PNR), statistical analysis and interpretation.
C. J. Kinane - Manuscript support and editing, experimental planning and data acquisition (PNR)
N. Satchell - experimental data acquisition (PNR).
G. Burnell - Project supervision, manuscript support and editing, experimental planning and supervision.
S. Langridge - Project supervision,  manuscript support and editing, experimental planning and supervision.
C. H. Marrows - Project conceptualization, supervision, manuscript original draft (lead), experimental planning and supervision.



\appendix

\section{Polarized Neutron Reflectivity}

\subsection{Structural Nuclear SLD fit\label{sec:Appendix_PNR_struc_fit}} 

\begin{figure*}[tbph]
    \includegraphics[width=\textwidth]{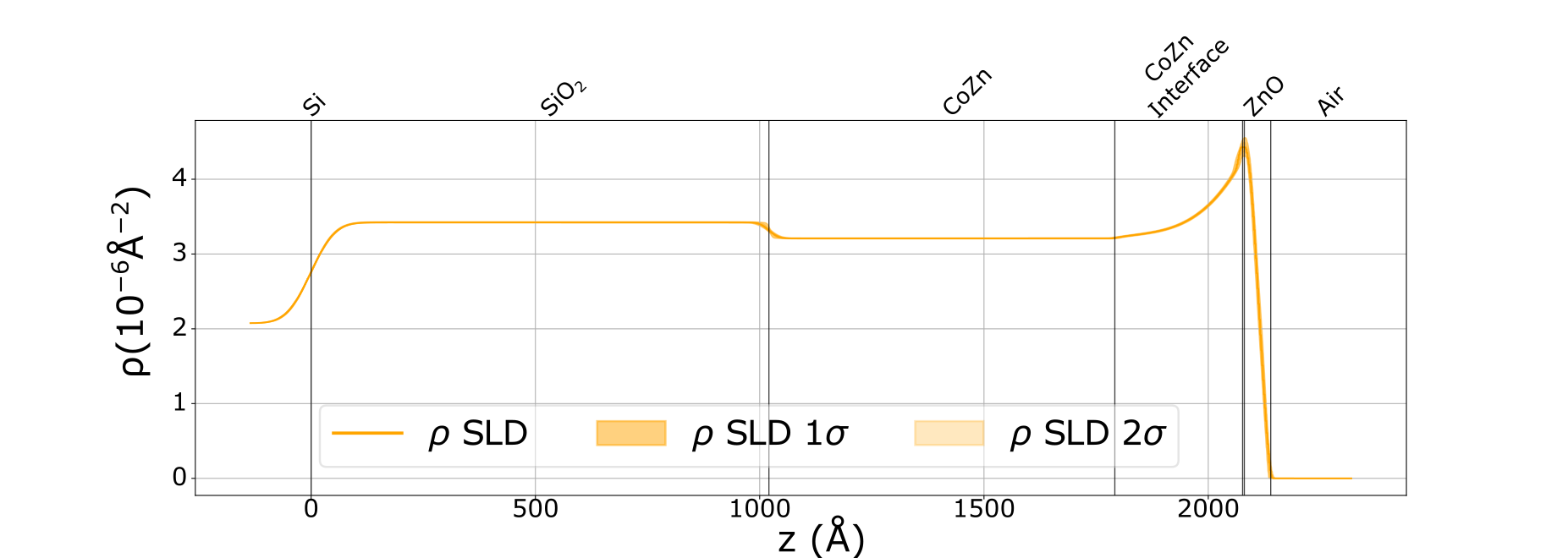}
    \caption{Structural nSLD profile derived from multi-fitting the 106~\degree C isotherm and the 25~\degree C 700~mT saturated data. The magnetic SLD (not shown here for clarity), of each data set was fitted using a freeform spline curve so as not to bias the nuclear fit by any preconceptions of the magnetic profile. The freeform spline was given enough freedom such that both helix or uniform profiles could be produced from one set of fitted parameters. This structural profile was then fixed for all of the magnetic profile fitting shown in sections \ref{sec:Polarized_neutron_reflectivity} and \ref{sec:Bayes_PNR}.
    \label{fig:Appendix_nSLD}}
\end{figure*}

The structural nuclear scattering length density (nSLD) of the CoZn sample used in this study is shown in Fig.~\ref{fig:Appendix_nSLD}. The structural fit was obtained prior to the specific mSLD model selection analysis described in sections \ref{sec:Polarized_neutron_reflectivity} and \ref{sec:Bayes_PNR}. This was to ensure that the only variable we were comparing between models was the magnetic profile. 

Since the structure was assumed to be invariant with respect to temperature, multiple data sets could be used in a large co-fit, with the aim to reduce the ambiguity in determining the nSLD. The structural parameters were shared between all datasets which were included in the co-fit, while the magnetic profile was unique for each dataset. The 106~\degree C isotherm in addition to the 25~\degree C 700~mT saturated data set, were used to determine the nSLD. While in principle more datasets from the study could have been used in this co-fit, practically the small improvement in uncertainty was outweighed by the large increase in convergence time when trying to include more data - indeed, the full co-fit never converged after significant numbers of iterations due to the very slow convergence rate.

As this co-fit was conducted before the final magnetic fitting, it was important to model the individual magnetic profiles for each of the datasets with enough freedom so that the structural fit was not skewed by any bias of what the magnetic profile should be. Thus, a freeform spline was chosen to model each magnetic profile. The spline comprised 3-control points which were spaced equidistant in $z$ across the CoZn layer - only the magnitude of each point in mSLD were fitted. This provided enough freedom to the model such that the spline could mimic a helical or uniform profile -- effectively removing the bias of the magnetism on the structural fit -- while minimizing model complexity such that the fit would still be able to converge.

Several models of varying complexity were tried to fit the nSLD, but due to the inter-diffuse nature of the CoZn/ZnO interface, the most consistent model to the data comprised spline interfaces between the CoZn and ZnO layers, and at the ZnO/air interface. These interfaces could not be satisfactorily modeled with the standard symmetric error function interface profiles. The rest of the profile was modeled with a traditional layered box model - regions of constant SLD interfaced with symmetric error function profiles.

When taking into account the different scattering contrasts between X-rays (see Fig.\ref{fig:xrs} panel (c)) and Neutrons, the structural SLD profiles broadly agree. The CoZn thickness is consistent between the two probes (where the CoZn layer and top interface combined total thickness for Neutrons is ${\sim}$1054~\AA\ and X-rays is ${\sim}$1020~\AA) and the overall interface widths of the top and bottom interfaces of the CoZn layer also follow a similar profile. The SiO$_2$-CoZn interface is sharp in both cases, while the CoZn-ZnO-Air region in both profiles is broad. Due to the improved contrast between CoZn and ZnO, the Neutrons provide a more nuanced picture of the surface interface that X-rays struggle to resolve. 

\subsection{Uniform Magnetic Profile\label{sec:Appendix_unifrom_mag_profile}} 

\begin{figure*}[tbph]
    \includegraphics[width=12.5cm]{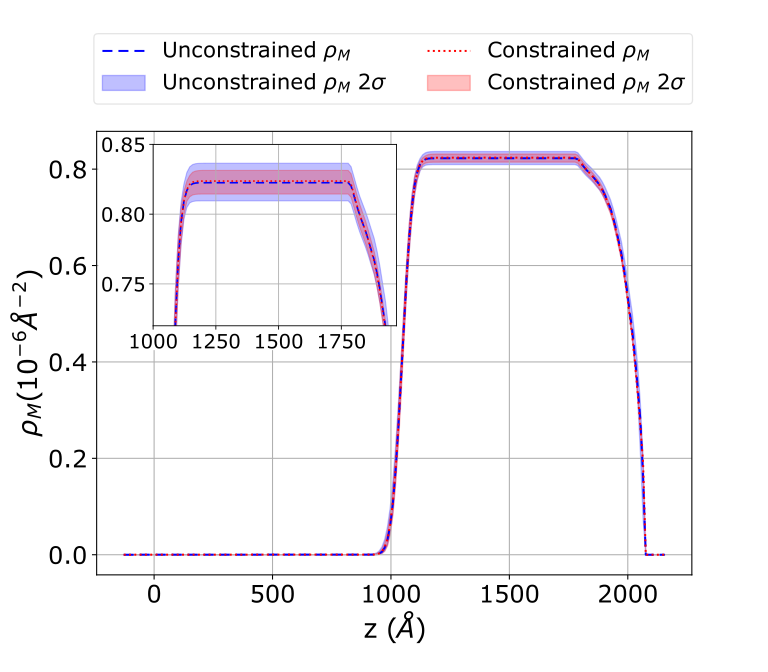}
    \caption{mSLD profile comparisons for the saturated 25~\degree C 700~mT data set. The upper interface of the mSLD profiles shown is determined by a 2-point spline. In the unconstrained profile the z-position of the control points was allowed to vary, while in the constrained profile the z-positions were matched to the structural z-positions of the CoZn interface. The 2-profiles match within 2-sigma justifying the use of the constrained profile to model the uniform magnetic profile. The unconstrained profile has slightly larger uncertainties, which is to be expected due to the increased freedom of the model.
    \label{fig:Appendix_uniform_mSLD}}
\end{figure*}

Due to the complexity of the CoZn/ZnO structural interface, modeling the uniform magnetization profile also required the use of an interface spline. Since freeform splines have significant freedom in terms of the profiles they can produce, this raised an issue when unconstrained at low fields, since the spline profile tried to mimic the helix profile. In order for a clear comparison to be made between the two profiles, the uniform profile was constrained such that the z positions of the spline were pinned to the same values of the structural interface. To justify this choice, we compared the fit of the unconstrained spline and the constrained spline for the saturated 25~\degree C 700~mT data set as shown in Fig.~\ref{fig:Appendix_uniform_mSLD}. The fits are almost identical, agreeing within 2-sigma (shown), and even within 1-sigma (not shown for clarity). The increased uncertainty shown in Fig.~\ref{fig:Appendix_uniform_mSLD} for the unconstrained profile, is to be expected since there will be more profiles that satisfy a similar likelihood (goodness of fit). From this, it was assumed that a uniform profile would follow the saturated profile shape, just having a smaller magnitude at lower applied fields. The constrained uniform profile was used for all of the analysis shown in sections \ref{sec:Polarized_neutron_reflectivity} and \ref{sec:Bayes_PNR}.

\bibliography{CoZn_PNR_biblography}

\end{document}